\documentclass[useAMS,referee]{biom}
\let\bibhang\undefined

\usepackage{amsmath, amsfonts, bm, threeparttable, subfigure, graphicx, float, natbib, makecell,color}

\newcommand{\sumi}{\sum_{i=1}^n}
\newcommand{\sumii}{\sum_{l=1}^n}
\newcommand{\sumj}{\sum_{j=1}^m}
\newcommand{\bbeta}{\bm \beta}

\newcommand{\balpha}{\bm \alpha}
\newcommand{\halpha}{\widehat{\bm \alpha}}

\newcommand{\hmu}{\widehat\mu}
\newcommand{\blambda}{\bm \lambda}
\newcommand{\hlambda}{\widehat{\bm \lambda}}

\newcommand{\hp}{\widehat p}

\newcommand{\bmeta}{\bm \eta}
\newcommand{\heta}{\widehat {\bm \eta}}
\newcommand{\bphi}{\bm \phi}
\newcommand{\hphi}{\widehat{\bm\phi}}
\newcommand{\bPhi}{\bm\Phi}
\newcommand{\bh}{\bm h}
\newcommand{\bX}{\bm X}
\newcommand{\bx}{\bm x}
\newcommand{\bv}{\bm v}
\newcommand{\hSigma}{\widehat \Sigma}
\newcommand{\mX}{\mathcal X}
\newcommand{\mY}{\mathcal Y}

\title[Transportability with summary statistics]{Transportable inference using target population summary statistics under covariate shift}

\author{Ying Sheng$^{1}$, Yifei Sun$^{2}$, and Chiung-Yu Huang$^{3,*}$ \email{ChiungYu.Huang@ucsf.edu} \\
$^{1}$State Key Laboratory of Mathematical Sciences, Academy of Mathematics and Systems Science, \\ Chinese Academy of Sciences, Beijing 100190, China.\\
$^{2}$Department of Biostatistics, Mailman School of Public Health, \\ Columbia University, New York, NY 10032, United States\\
$^{3}$Department of Epidemiology \& Biostatistics,  University of California at San Francisco, \\ San Francisco, CA 94158, United States}

\begin{document}






\label{firstpage}


\begin{abstract}
Transporting findings from a study population to a target population is central to evidence-based decision-making in real-world settings. Most existing methods require individual-level data from both populations to account for covariate shift. However, privacy regulations and data-sharing constraints often preclude access to such data from the target population, leaving only covariate summaries available for analysis. In this paper, we develop transportability methods that enable valid inference using source individual-level data and target covariate summaries. First, we apply entropy balancing to transportability, enabling source individual-level data to be adjusted to match the target covariate moments. We establish asymptotic normality for the entropy balancing estimator and propose a variance estimator to account for uncertainty in covariate summaries. Second, we develop a new transportability method that allows more flexible modeling of covariate shift, thereby accounting for covariate shift and uncertainty in covariate summaries simultaneously. Asymptotic normality for the proposed estimator is established and its asymptotic variance is consistently estimated. The proposed method offers greater flexibility in accounting for covariate shift and thus permits consistent estimation and valid inference when the validity of entropy balancing is not guaranteed. The proposed methods are evaluated by simulations and illustrated with an analysis of Surveillance, Epidemiology, and End Results breast cancer data. 
\end{abstract}




\maketitle

\section{Introduction}

Transportability, which refers to extending statistical findings from a source population to a target population of interest, has drawn increasing attention in recent years. Transportability methods have been successfully applied in various fields, particularly when data collection from the target population is time-consuming or expensive. Most transportability methods focus on accounting for distributional shift between the source and target populations, which arises from differences in study designs, subject characteristics and so on \citep{degtiar2023review, colnet2024causal}. In general, existing methods require access to individual-level data from both the source and target populations \citep{rudolph2017robust, dahabreh2019generalizing, dahabreh2020extending, josey2022calibration, dahabreh2023efficient, wu2023transfer}. However, in real-world applications such as healthcare and finance, individual-level data from the target population are often unavailable due to privacy regulations and data-sharing constraints. 
In such settings, the aforementioned methods cannot be applied directly.

In this paper, we are interested in the setting where individual-level data from the source population are accessible, but only covariate summaries from the target population are available. To characterize distributional shift between the source and target populations, we focus on covariate shift, where the distributions of covariates differ across two populations but the conditional distribution of the outcome given covariates remains the same. Covariate shift, which arises from differences in subject characteristics, is a commonly used assumption. To account for heterogeneity in the distributions of covariate, methods using target summary statistics have been studied \citep{josey2021transporting, zheng2022risk, chen2023entropy, chu2023targeted, chattopadhyay2024one}.
When uncertainty in the target summary statistics is not ignorable, \cite{chen2026confidence} proposed a resampling-based perturbation approach for confidence interval construction. They resampled the source individual-level data and perturbed target summary statistics based on the empirical covariance matrix from the target population. 
When only the diagonal elements of the empirical covariance matrix were available, they estimated the corresponding correlation matrix using the source individual-level data. However, the consistency of the resulting covariance matrix estimator is not guaranteed under covariate shift.
To fill this gap, our goal is to conduct inferences about the target population with theoretical justification by accounting for covariate shift and uncertainty in the target covariate summaries simultaneously.

To account for covariate shift, one common method is to construct weights such that the reweighted source covariate distribution approximates the target covariate distribution \citep{hainmueller2012entropy, chan2016globally, zhao2017entropy, lee2023improving}. Entropy balancing (EB) \citep{hainmueller2012entropy} is a popular reweighting method, which was originally proposed to adjust the control group to match the prespecified covariate moments in the treated group. EB naturally extends to transportability when only covariate moments from the target population are accessible. Specifically, we apply EB to calculate weights such that the reweighted source sample moments match the target sample moments, and then obtain a weighted estimator for the target parameter of interest. Under regularity conditions, we establish asymptotic normality for the EB estimator. 
Estimating the asymptotic variance of the EB estimator requires a consistent estimator for the asymptotic covariance matrix of the target covariate summaries. To address this challenge, we propose a weighted estimator for this covariance matrix that relies solely on the source individual-level data and the target covariate summaries, thereby obtaining a variance estimator.
Both the EB estimator and the variance estimator are consistent if $\log \{w(\bx)\}$ is linear in $\bPhi(\bx)$, where $w(\bx)$ is the density ratio of the target distribution to the source distribution, and $\bPhi(\bx)$ is a given vector function of covariates whose target sample moments are available.

When only a limited set of target covariate summaries is available, the resulting $\bPhi(\bx)$ may not involve important nonlinearities or interactions. As a result, $\log \{w(\bx)\}$ cannot be well approximated by a linear combination of $\bPhi(\bx)$ and EB becomes restrictive in accounting for covariate shift. To remove this restriction, we develop a new transportability method that allows more flexible modeling of $w(\bx)$ using a  model $\pi(\bx;\balpha)$. Model-based reweighting methods have been studied in the literature \citep{ qin2008efficient, han2013estimation, imai2014covariate, chan2014oracle, han2014multiply}. However, these existing methods cannot be applied directly to estimate $\balpha$ when individual-level covariate data are not available from the target population. 
To tackle this problem, we propose to estimate $\balpha$ by minimizing an augmented objective function subject to constraints derived from target summary statistics, where an augmentation is incorporated to account for uncertainty in the target summary statistics. The proposed estimator is shown to be consistent and asymptotically normal when $w(\bx) $ is correctly specified by $\pi(\bx;\balpha)$. To evaluate the validity of this specification, we construct a test statistic and establish its asymptotic chi-squared distribution under the null hypothesis. Compared with the EB method, the proposed method offers greater flexibility in accounting for covariate shift. Specifically, the EB method becomes a special case of the proposed method by specifying $\log\{\pi(\bx;\balpha)\}$ to be a linear combination of $\bPhi(\bx)$. When $\log\{w(\bx)\}$ is linear in $\bPhi(\bx)$, the proposed estimator is asymptotically at least as efficient as the EB estimator. When $\log\{w(\bx)\}$ is not linear in $\bPhi(\bx)$,  the proposed method permits consistent estimation and valid inference by allowing more appropriate forms of $\pi(\bx;\balpha)$.

\section{Notation and framework} 
\label{setup}

Let $Y\in\mathbb{R}$ and $\bX \in\mathbb{R}^d$ denote the outcome and covariates in the source population, respectively.  Similarly, let $Y^*\in\mathbb{R}$ and $\bX^*\in\mathbb{R}^d$ represent the outcome and covariates in the target population. Denote by $f(y,\bx)$ and $f^*(y,\bx)$ the joint density functions of $(Y, \bX )$ and $(Y^*, \bX^*)$, respectively. Our goal is to leverage knowledge from the source population to conduct inferences about the target population. However, when the two populations differ in their underlying characteristics, naively transporting evidence from one to the other may result in biased or even misleading conclusions. To address this challenge, we adopt a commonly used framework that assumes covariate shift between the source and target populations. Specifically, assume that the joint distribution in the target population satisfies
\begin{eqnarray}
\label{x-shift}
f^*(y,\bx)=f(y,\bx) w(\bx),
\end{eqnarray}
where $w(\bx)$ is an unspecified covariate shift function that characterizes the heterogeneity in the marginal distributions of covariates between the source and target populations. 
Under this model, the marginal distributions of covariates may differ across the two populations, but the conditional distribution of the outcome given covariates remains the same.

Our goal is to estimate $\mu^*=\operatorname{E}^*(Y^*)$ in the target population,  where $\operatorname{E}^*(\cdot)$ denotes the expectation taken with respect to the target distribution. Suppose the source study consists of $n$ independent and identically distributed (i.i.d.) observations, denoted by $\{(y_i, \bx_i), i = 1,\ldots, n\}$. A natural approach is to exploit the invariance of the conditional outcome distribution under covariate shift, where $\operatorname{E}(Y | \bX=\bx)=\operatorname{E}^*(Y^* | \bX^*=\bx)$ for all $\bx$  with $\operatorname{E}(\cdot)$ denoting the expectation taken with respect to the source distribution. This leads to the regression-based estimator $m^{-1}\sumj \widehat{\operatorname{E}}(Y| \bX=\bx_j^*)$, where  $\widehat{\operatorname{E}}(Y| \bX)$ is trained on the source data $\{(y_i, \bx_i), i = 1, \ldots, n\}$ and applied to i.i.d. covariate data $\{\bx_j^*, j=1,\ldots,m\}$ from the target population. Alternatively, it follows from \eqref{x-shift} that $\mu^* = \operatorname{E}\{w(\bX)Y \}$.
Thus, one can construct a weighted estimator 
$
n^{-1} \sum_{i=1}^n \widehat{w}(\bx_i) y_i,
$
where $\widehat{w}(\bx)$ is a consistent estimator of $w(\bx)$. Existing methods, such as kernel mean matching  \citep{huang2006correcting} or least squares importance fitting \citep{kanamori2009least}, can be applied to estimate $w(\bx)$ directly. Moreover, by the equivalence between density ratio estimation and classification of source/target membership \citep{Prentice1979logistic}, one can estimate $w(\bx)$ by fitting a logistic regression or any well-calibrated classifier. Finally, combining the two estimators gives the augmented estimator $n^{-1} \sum_{i=1}^n \widehat{w}(\bx_i) \{ y_i - \widehat{\operatorname{E}}(Y| \bX= \bx_i) \} + m^{-1} \sum_{j=1}^m \widehat{\operatorname{E}}(Y| \bX=\bx_j^*)$, which is consistent for $\mu^*$ if either the covariate shift model or the outcome regression model is correctly specified.

In many practical applications, the target individual-level data are not available due to privacy concerns, regulatory restrictions, or logistical constraints. Instead, only aggregate covariate information is accessible, typically in the form of summary statistics.  In such cases, the aforementioned methods cannot be applied directly, as neither $\widehat{\operatorname{E}}(Y| \bX=\bx_j^*)$ nor $\widehat{w}(\bx)$  can be obtained without individual-level covariate data from the target population.

\section{Entropy balancing with summary statistics}
\label{method1}

Originally developed by \cite{hainmueller2012entropy} for causal inference, EB constructs weights to align covariate distributions between treated and control groups. This method naturally extends to transportability, enabling source data to be adjusted to match the target covariate characteristics. 
In this section, we apply EB to estimate  \(\mu^* \). Let $\bPhi(\bx)$ be a given vector function of covariates whose target sample moments  \(\widehat{\bphi}^* =  m^{-1} \sum_{j=1}^m  \bPhi(\bx_j^*)\) are available. Common choices of $\bPhi(\bx)$ include $\operatorname{I}(\bx\in \Omega)$ and $\bx$, where $\hphi^*$ estimates the proportion of $\bX^*$ in a subgroup $\Omega$ and mean of $\bX^*$ in the target population, respectively.

To estimate \(\mu^*\), we obtain the EB weights by minimizing $\sum_{i=1}^n np_i \log (n p_i)$ subject to $p_i > 0$, $\sum_{i=1}^n p_i = 1$ and $\sum_{i=1}^n p_i \{ \bPhi(\bx_i) - \widehat{\bphi}^* \} = \bm 0$, where the last constraint matches the reweighted source sample moments to the target sample moments $\widehat{\bphi}^*$.
The entropy loss $\sum_{i=1}^n np_i \log (n p_i)$ quantifies the discrepancy between $(p_1,\ldots,p_n)$ and $(n^{-1},\ldots,n^{-1})$, corresponding to the limiting case of the Cressie and Read family of discrepancies $\sum_{i=1}^n \{(np_i)^{\gamma+1}-1\}/\{\gamma(\gamma+1)\}$ as $\gamma \to 0$ \citep{cressie1984multinomial}. The entropy loss is adopted because it is robust under misspecification and constrains the weights to be non-negative \citep{Imbens1998Information,hainmueller2012entropy}. Applying the standard Lagrange multiplier argument yields
\begin{eqnarray}
\label{pi}
\widehat p_i = \frac{\exp[
\hlambda^\top \{\bPhi(\bx_i)-\hphi^* \}]}{\sumii \exp[
\hlambda^\top \{\bPhi(\bx_l)-\hphi^* \}] }, \; i=1,\ldots,n,
\end{eqnarray}
where $\hlambda=\arg\max_{\blambda} (-n\log \sumi \exp[\blambda^\top \{\bPhi(\bx_{i})-\hphi^*\}] ),$
or, equivalently, $\hlambda$ is the solution to 
$
\sumi  \{\bPhi(\bx_i)-\hphi^*  \} \exp [\blambda^\top  \{\bPhi(\bx_i)-\hphi^*  \} ]= \bm 0.
$
The EB estimator for $\mu^*$ is 
$
\hmu_{\rm EB} = \sumi \hp_i y_i.
$

As noted in \cite{zhao2017entropy}, the EB method enjoys a double robustness property in the sense that $\widehat{\mu}_{\rm EB}$ is consistent for $\mu^*$ if either a linear model for the outcome  $\operatorname{E}(Y|\bX)= \beta_0+\bbeta^\top\bPhi(\bX)$ or a log-linear model for covariate shift $\log\{w(\bx)\}=  a_0+\bm a^\top\bPhi(\bx)$ is correctly specified. To establish consistency of $\hmu_{\rm EB}$, we note that $\hmu_{\rm EB} = \sumi \hp_i y_i$ converges to $\mu_{\rm EB} = \operatorname{E}\{ \pi_{\rm EB}(\bX) Y\}$ as $m,n\to \infty$, where 
$
\pi_{\rm EB}(\bx) =   \exp[\blambda_{0}^\top \{\bPhi(\bx)-\bphi^*\}]/\operatorname{E} ( \exp[\blambda_{0}^\top \{\bPhi(\bX)-\bphi^*\}]  )
$
with $\bphi^*=\operatorname{E}^*\{\bPhi(\bX^*)\}$ and $\blambda_0$ satisfying
\begin{eqnarray}
\label{eq-lambda}
\operatorname{E}\left(  \left\{  \bPhi(\bX)-\bphi^*  \right\} \exp[\blambda_0^\top  \{\bPhi(\bX)-\bphi^*  \} ]\right)= \bm 0.
\end{eqnarray}
If $\operatorname{E}(Y|\bX)= \beta_0+\bbeta^\top\bPhi(\bX)$, we have $\mu_{\rm EB} = \operatorname{E}  [\pi_{\rm EB}(\bX) \{\beta_0+\bbeta^\top\bPhi(\bX)\} ] = \beta_0+ \bbeta^\top\bphi^*= \mu^*,$ 
where $\operatorname{E}\{\pi_{\rm EB}(\bX) \bPhi(\bX)\}=\bphi^*$ follows from \eqref{eq-lambda}. On the other hand, if $\log\{w(\bx)\}=  a_0+\bm a^\top\bPhi(\bx)$, we can show that $w(\bx)=\pi_{\rm EB}(\bx)$, leading to 
$
\mu_{\rm EB}=\operatorname{E}\{\pi_{\rm EB}(\bX) Y\}=\operatorname{E}\{w(\bX) Y\}=\mu^*.
$
The proof of double robustness is given in Section 2 of the supplementary materials. The large-sample property of $\widehat\mu_{\rm EB}$ is summarized in Theorem \ref{thm1}, with the proof given in Section 3 of the supplementary materials.

\begin{theorem}  
\label{thm1}
Under conditions (C1)--(C5) in Section 1 of the supplementary materials, as $n,m\to \infty$, we have $\sqrt{n}(\hmu_{\rm EB}-\mu_{\rm EB})$ converges in distribution to a normal distribution with mean zero and variance $\sigma^2_{\rm EB} = \operatorname{var}\{U(\bX,Y)\}+\rho\bm \omega^\top \Sigma^* \bm \omega,$ where  
$U(\bx,y)= \pi_{\rm EB}(\bx) (y  -  \mu_{\rm EB}) -  \pi_{\rm EB}(\bx) \bm \omega^\top \{\bPhi(\bx)-\bphi^*\},$ 
$\rho = \lim_{m,n\to \infty} n m^{-1} \in (0,\infty)$, $\Sigma^* =  \operatorname{E}^*  [ \{ \bPhi(\bX^*)-\bphi^* \}^{\otimes 2}]$ and $\bm \omega =  \left( \operatorname{E} [\pi_{\rm EB}(\bX) \{\bPhi(\bX)-\bphi^*\}^{\otimes 2}] \right)^{-1} \operatorname{E} [Y\pi_{\rm EB}(\bX) \{\bPhi(\bX)-\bphi^*\}] $ with $\bm a^{\otimes 2}=\bm a\,\bm a^\top$ for a vector $\bm a$.
\end{theorem}

Conditions (C1)--(C5) require that the true parameters lie in a compact set, the estimating equation \eqref{eq-lambda} has a unique solution, the function $\exp[\blambda^\top \{\bPhi(\bx)-\bphi\}]$ is smooth with bounded derivatives with respect to $(\blambda^\top,\bphi^\top)^\top$, $\Sigma^*$ is positive definite, and the participation probability is positive.
Theorem \ref{thm1} shows that $\sigma_{\rm EB}^2$ depends on $\Sigma^*$, which is the asymptotic covariance matrix of $\sqrt{m}(\hphi^*-\bphi^*)$, so we propose a weighted estimator for $\Sigma^*$ only using the source individual-level data and $\widehat{\bphi}^*$. 
It follows from \eqref{x-shift} that
$\Sigma^*=\operatorname{E} \{w(\bX)\bPhi(\bX)^{\otimes 2}\} - \bphi^{*\otimes 2} $.
Because $\hphi^*$ is consistent for $\bphi^*$ and EB reweights the source data to match the target covariate distribution via $\widehat p_i$ defined by \eqref{pi}, we propose to estimate $\Sigma^*$ by 
$
\widehat \Sigma_{\rm EB} = \sumi \widehat p_i \bPhi(\bx_i)^{\otimes 2} -\hphi^{*\otimes 2}.
$
It can be shown that $\widehat \Sigma_{\rm EB}$ converges to $\operatorname{E} \{\pi_{\rm EB}(\bX)\bPhi(\bX)^{\otimes 2}\} - \bphi^{*\otimes 2}$ as $m,n\to \infty$. If $\log \{ w (\bx)\}$ is linear in $\bPhi(\bx)$,  we can show that $\pi_{\rm EB}(\bx)=w(\bx)$ and thus $\widehat \Sigma_{\rm EB}$ is consistent for $\Sigma^*$.  Moreover, $\operatorname{var}\{U(\bX,Y)\}$ and $\bm \omega$ can be consistently estimated by replacing the expectations with empirical counterparts and $\bphi^*$ with $\hphi^*$, 
denoted by $\widehat \sigma_{U}^2$ and $\widehat {\bm \omega}$, respectively. Then $\sigma_{\rm EB}^2$ can be estimated by
$
\widehat\sigma_{\rm EB}^2 = \widehat \sigma_{U}^2 + nm^{-1} \widehat {\bm \omega}^\top \widehat \Sigma_{\rm EB} \widehat {\bm \omega}.
$
If $\log \{ w (\bx)\}$ is linear in $\bPhi(\bx)$, both $\hmu_{\rm EB}$ and $\widehat\sigma_{\rm EB}^2$ are consistent and thus EB is suitable for estimation and inference of $\mu^*$. More details for obtaining $\widehat\sigma_{\rm EB}^2$ are given in Section 3 of the supplementary materials.

\vspace{-0.2 in}

\section{Transportability via model-based reweighting}
\label{method2}

It is worthwhile to point out that EB was originally developed for settings where individual-level covariate data from the target population are available. In that case, target moments can be computed directly to permit an arbitrary choice of matched moments and, consequently, outcome and covariate shift models that include any important covariates and their transformations or interactions. However, when only aggregate summaries are released, EB becomes restrictive in accounting for covariate shift. The fitted covariate shift model is effectively constrained by the limited set of reported moments, which fixes the functional form spanned by $\bPhi(\bx)$ and curtails flexibility to capture important nonlinearities or interactions. 
To remove this restriction, we propose a new transportability method that allows more flexible modeling of $w(\bx)$ while matching target moments via minimum divergence calibration. This added flexibility permits consistent estimation and valid inference when $\log\{w(\bx)\}$ is not linear in $\bPhi(\bx)$, a setting in which the validity of EB is not guaranteed.

We postulate a working model $\pi(\bx;\balpha)$ for $w(\bx)$, where the form of $\pi(\bx;\balpha)$ is specified and $\balpha \in \mathbb{R}^{d_{\balpha}}$ is the unknown parameter.  Compared with EB, the proposed method offers greater flexibility in modeling $w(\bx)$ using $\pi(\bx;\balpha)$ because $\log \{\pi(\bx;\balpha)\}$ is not necessarily linear in $\bPhi(\bx)$ and can capture complex dependencies on $\bx$.   
If $w(\bx)=\pi(\bx;\balpha_0)$ for some $\balpha_0 $, we have $\mu^*= \int_{\mX}\int_{\mY} \pi(\bx;\balpha_0)y d F(y,\bx)$, where $\mX$ is the support of $\bX$, $\mY$ is the support of $Y$, and $F(y,\bx)$ is the cumulative distribution function (CDF) of $(Y,\bX)$. Therefore, we focus on estimating $\balpha_0$ for estimation and inference of $\mu^*$. 
By $w(\bx)=\pi(\bx;\balpha_0)$, we can summarize aggregate covariate information from the target population via $K+1$ estimating equations 
\begin{eqnarray}
\label{EE}
\int_{\mX}\int_{\mY}  \{\pi(\bx;\balpha_0)-1\} {\rm d} F(y,\bx) =0,\;\;
\int_{\mX}\int_{\mY}  \{\bPhi(\bx)-\bphi^*\}\pi(\bx;\balpha_0) {\rm d} F(y,\bx) =\bm 0,
\end{eqnarray}
where $K$ is the dimension of $\bphi^*$. For model identifiability, we assume that $d_{\balpha}\le K+1$.  
For $i=1,\ldots,n$, let $q_i={\rm d} F(y_i,\bx_i)$. 
When the population parameter $\bphi^*$ is available, the exponential tilting estimator \citep{kitamura1997information,  Imbens1998Information} for $\balpha_0$ can be obtained by minimizing $\sumi n q_i\log (nq_i)$ subject to the constraints 
\begin{eqnarray}
\label{constraints}
q_i>0,\;\;\sumi q_i=1,\;\; \sumi q_i \{\pi(\bx_i;\balpha)-1\}=0,\;\; \sumi q_i \pi(\bx_i;\balpha) \{\bPhi(\bx_i)-\bphi^*\}=\bm 0,
\end{eqnarray} 
where the first two constraints in \eqref{constraints} ensure that $F(y,\bx)$ is a proper CDF and the last two constraints in \eqref{constraints} are obtained by approximating the population estimating equations in \eqref{EE} using the source individual-level data.

In practice, the target covariate summaries $\hphi^*$, rather than $\bphi^*$, are often available. We propose to account for uncertainty in $\hphi^*$ by treating $\hphi^*$ as the realized value of a random vector \citep{zhang2020generalized}. Note that $\sqrt{m}(\hphi^*-\bphi^*)$  converges in distribution to a multivariate normal distribution with mean zero and covariance matrix $\Sigma^*$ as $m\to \infty$. We can obtain the asymptotic log-likelihood up to a constant, that is, $-m (\hphi^*-\bphi)^\top \Sigma^{*-1}(\hphi^*-\bphi)/2 $. Since $\Sigma^*$ is unknown, we propose to estimate $\balpha_0$ and $\bphi^*$ by minimizing 
$
\sumi n q_i\log (nq_i) + m (\hphi^*-\bphi)^\top V^{-1} (\hphi^*-\bphi)/2
$ 
subject to $q_i>0$, $\sumi q_i=1$, $\sumi q_i \{\pi(\bx_i;\balpha)-1\}=0$ and $\sumi q_i \pi(\bx_i;\balpha) \{\bPhi(\bx_i)-\bphi\}=\bm 0,$
where $V$ is any given positive definite matrix of dimension $K$. 
To solve the constrained minimization problem, consider the Lagrangian function
$L_V = \sumi nq_i\log(n q_i) +  m (\hphi^*-\bphi)^\top V^{-1} (\hphi^*-\bphi)/2- n\eta_0\left( \sumi q_i-1\right) - n\bmeta^\top \sumi q_i \bh(\bx_i;\balpha,\bphi), $
where $\bh(\bx;\balpha,\bphi)=(\pi(\bx;\balpha)-1, \pi(\bx;\balpha) \{\bPhi(\bx)-\bphi\}^\top)^\top$, $\eta_0$ and $\bmeta=(\eta_1,\ldots,\eta_{K+1})^\top$ are Lagrange multipliers. 
Applying the standard Lagrange multiplier argument yields
\begin{eqnarray}
\label{qi} 
q_i(\bmeta,\balpha,\bphi) = \frac{\exp\{\bmeta^\top\bh(\bx_i;\balpha,\bphi)\}}{\sumii \exp\{\bmeta^\top\bh(\bx_{l};\balpha,\bphi)\}}, \;\;i=1,\ldots,n,
\end{eqnarray}
where $\bmeta$ satisfies
$
\sumi \exp\{\bmeta^\top\bh(\bx_i;\balpha,\bphi)\}\bh(\bx_i;\balpha,\bphi)=\bm 0
$
due to $\sumi q_i(\bmeta,\balpha,\bphi) \bh(\bx_i;\balpha,\bphi)=\bm 0$. 
Substituting $q_i(\bmeta,\balpha,\bphi)$ back into the Lagrangian function yields, up to a constant,
\begin{eqnarray}
\label{ellV}
\ell_V(\bmeta,\balpha,\bphi) =  - n\log  \sumi \exp\{\bmeta^\top \bh(\bx_i;\balpha,\bphi)\}   +  \frac{m}{2} (\hphi^*-\bphi)^\top V^{-1} (\hphi^*-\bphi). 
\end{eqnarray} 
Following the argument in \cite{Imbens1998Information}, the constrained minimization problem can be written as a saddle point problem 
$
\min_{\balpha,\bphi}\max_{\bmeta} \ell_V(\bmeta,\balpha,\bphi),
$
with the resulting estimators denoted by $(\heta_V,\halpha_V,\hphi_V)$.
Due to $\mu^*=\int_{\mX}\int_{\mY} \pi(\bx;\balpha_0)y {\rm d} F(y,\bx)$, we can estimate $\mu^*$ by 
$
\hmu_V =\sumi q_i(\heta_V,\halpha_V,\hphi_V) \pi(\bx_i;\halpha_V)y_i,  
$
where $q_i(\bmeta,\balpha,\bphi)$, $i=1,\ldots,n$, are defined by \eqref{qi}.

We have proved in Section 4 of the supplementary materials that if $w(\bx)=\pi(\bx;\balpha_0)$, $\hmu_V$ is consistent for any positive definite matrix $V$  and choosing $V$ as a consistent estimator of $\Sigma^*$ yields the most efficient estimator. This motivates us to propose a consistent estimator for $\Sigma^*$. It follows from \eqref{x-shift} and $w(\bx)=\pi(\bx;\balpha_0)$ that $\Sigma^* = \operatorname{E} \{ \pi(\bX;\balpha_0) \bPhi(\bX)^{\otimes 2} \} - \bphi^{*\otimes 2} $. 
By \eqref{EE}, a consistent initial estimator $\widehat\balpha^{(0)}$ can be obtained by minimizing $\|n^{-1}\sumi \bh(\bx_i;\balpha,\hphi^*)\|^2$. Therefore, a consistent initial estimator of $\Sigma^*$ is
$
\hSigma^{(0)} = n^{-1} \sumi  \pi(\bx_i;\widehat\balpha^{(0)})\bPhi(\bx_i)^{\otimes 2}- \hphi^{* \otimes 2}.  
$
To avoid confusion, let $(\heta,\halpha,\hphi)$ be the estimators obtained by solving 
$
\min_{\balpha,\bphi}\max_{\bmeta} \ell(\bmeta,\balpha,\bphi),
$
where $\ell(\bmeta,\balpha,\bphi)$ is calculated by replacing $V$ with $\hSigma^{(0)}$ in $\ell_V(\bmeta,\balpha,\bphi)$ defined by \eqref{ellV}. Thus, the proposed efficient estimator for $\mu^*$ is 
$
\hmu=\sumi \widehat q_i \pi(\bx_i;\halpha)y_i,
$
where $\widehat q_i=q_i(\heta,\halpha,\hphi)$ with $q_i(\bmeta,\balpha,\bphi)$ defined by \eqref{qi}. 
The large-sample property of $\hmu$ is summarized in Theorem \ref{thm2}, with the proof given in Section 4 of the supplementary materials.

\begin{theorem}
\label{thm2} 
Assume that there exists some $\balpha_0\in \mathbb{R}^{d_{\balpha}}$ such that $w(\bx)=\pi(\bx;\balpha_0)$.  Under conditions (C4)--(C9) in Section 1 of the supplementary materials, as $ n,m\to \infty$, we have $\sqrt{n}(\hmu-\mu^*)$ converges in distribution to a normal distribution with mean zero and variance 
$$
\sigma^2 = \operatorname{var}\{w(\bX)Y\} - \bv^\top  \begin{pmatrix}
 \Sigma_{\bh} + \rho \mathcal{J}_{\bphi} \Sigma^* \mathcal{J}_{\bphi}^\top &  \mathcal{J}_{\balpha} \\
  \mathcal{J}_{\balpha}^\top & \bm 0_{d_{\balpha} \times d_{\balpha}}
\end{pmatrix}^{-1} \bv,
$$
where $\Sigma_{\bh}=\operatorname{E}\{\bh(\bX;\balpha_0,\bphi^*)^{\otimes 2}\}$,  $\Sigma^* =  \operatorname{E}^* [ \{ \bPhi(\bX^*)-\bphi^*\}^{\otimes 2} ]$, $\rho = \lim_{m,n\to \infty} n m^{-1} \in (0,\infty)$, $\mathcal{J}_{\bphi} = \operatorname{E}\{ \partial \bh(\bX;\balpha,\bphi)/\partial \bphi |_{\balpha=\balpha_0,\bphi=\bphi^*}\}$,
$\mathcal{J}_{\balpha} = \operatorname{E}\{ \partial \bh(\bX;\balpha,\bphi)/\partial \balpha |_{\balpha=\balpha_0,\bphi=\bphi^*}\}$ and 
$$
\bv = \begin{pmatrix}
 \operatorname{E}\{Y w(\bX)\bh(\bX;\balpha_0,\bphi^*)\}\\
 \operatorname{E}\left\{Y \frac{\partial \pi(\bX;\balpha)}{\partial\balpha}|_{\balpha=\balpha_0}\right\}
\end{pmatrix}.
$$
\end{theorem}

Conditions (C6)--(C9) require that the true parameters lie in a compact set, the estimating equation $\operatorname{E}\{\bh(\bX;\balpha,\bphi)\}=\bm 0$ has a unique solution, $\Sigma_{\bh}$ is positive definite,  and $\bh(\bX;\balpha,\bphi)$ is smooth with bounded derivatives with respect to $(\balpha^\top,\bphi^\top)^\top$.
Theorem \ref{thm2} shows that the proposed method can consistently estimate $\mu^*$ and properly account for the uncertainty of $\hphi^*$ when $w(\bx)$ is correctly specified by $\pi(\bx;\balpha)$. We propose a consistent estimator for $\sigma^2$ as follows. Given $\widehat q_i$, $\halpha$ and $\hphi$, we update the estimation of $\Sigma^*$ by
$\hSigma= \sumi \widehat q_i  \pi(\bx_i;\widehat\balpha)\bPhi(\bx_i)^{\otimes 2}- \hphi^{\otimes 2}.$
Moreover, by replacing the expectations with their empirical counterparts and unknown parameters with their estimators, we can consistently estimate $\operatorname{var}\{w(\bX)Y\}$, $\Sigma_{\bh}$, $\mathcal{J}_{\bphi}$, $\mathcal{J}_{\balpha}$, and $\bv$, denoted by $\widehat \sigma^2_{w}$, $\widehat \Sigma_{\bh}$, $\widehat {\mathcal{J}}_{\bphi}$, $\widehat {\mathcal{J}}_{\balpha}$, and $\widehat {\bv}$, respectively. Therefore, we can obtain a plug-in estimator for $\sigma^2$, with the details given in Section 4 of the supplementary materials.

The consistency of $\hmu$ relies on the correct specification of $\pi(\bx;\balpha)$. In practice, $\pi(\bx;\balpha)$ can be specified by comparing prior information regarding covariate shift between the source and target populations, including study designs, inclusion and exclusion criteria, and available summary statistics. Next, we propose a model checking procedure to check the validity of $\pi(\bx;\balpha)$. Specifically, consider the null hypothesis $\mathcal H_0:$ $w(\bx)=\pi(\bx;\balpha_0)$ for some $\balpha_0  \in \mathbb{R}^{d_{\balpha}}$. 
Under $\mathcal H_0$, we have $\operatorname{E}\{\bh(\bX;\balpha_0,\bphi^*)\}=\bm 0$, where $\bh(\bx;\balpha,\bphi)$ is of dimension $K+1$. To test $\mathcal H_0$, consider checking $\operatorname{E}\{\bh(\bX;\balpha_0,\bphi^*)\}=\bm 0$ based on  $\bh_n(\halpha,\hphi^*)$, where $\bh_n(\balpha,\bphi)= n^{-1} \sumi \bh(\bx_i;\balpha,\bphi)$.
Under the conditions in Theorem \ref{thm2}, we can prove that 
$\sqrt{n} \bh_n(\halpha,\hphi^*) = W_\rho^{1/2} \mathcal{P} W_\rho^{-1/2} \sqrt{n}\{ \bh_n(\balpha_0,\bphi^*) +  \mathcal J_{\bphi}  (\hphi^*-\bphi^*) \} + o_p(1),$
where $\mathcal{P}$ is an idempotent matrix with rank $K+1-d_{\balpha}$, and $W_\rho = \Sigma_{\bh} + \rho   \mathcal J_{\bphi}  \Sigma^*  \mathcal J_{\bphi}^\top$ with $\Sigma_{\bh}$,  $\rho$, $\mathcal J_{\bphi}$ and $\Sigma^*$  defined in Theorem \ref{thm2}. Under $\mathcal H_0$, $ W_\rho^{-1/2}  \sqrt{n} \{ \bh_n(\balpha_0,\bphi^*) + \mathcal J_{\bphi} (\hphi^*-\bphi^*) \}$ converges in distribution to a standard multivariate normal variable as $n,m\to \infty$. This indicates that $n \bh_n(\halpha,\hphi^*)^\top W_\rho^{-1} \bh_n (\halpha,\hphi^*) $ is asymptotically chi-squared. Therefore, we consider the test statistic 
$\mathcal T =  n \bh_n(\halpha,\hphi^*)^\top \widehat W_\rho^{-1} \bh_n (\halpha,\hphi^*),$ where $\widehat W_\rho$ is a consistent estimator of $W_\rho$ under $\mathcal H_0$. Theorem \ref{thm3} summarizes the asymptotic distribution of $\mathcal T$ under $\mathcal H_0$, with the proof given in Section 5 of the supplementary materials. 

\begin{theorem}
\label{thm3}
Under conditions (C4)--(C9) in Section 1 of the supplementary materials and the null hypothesis $\mathcal H_0:$ $w(\bx)=\pi(\bx;\balpha_0)$ for some $\balpha_0  \in \mathbb{R}^{d_{\balpha}}$, the test statistic $\mathcal T$ converges in distribution to a $\chi^2$ distribution with $K+1-d_{\balpha}$ degrees of freedom as $n,m\to \infty$.
\end{theorem}

\section{ Comparison between the EB method and the proposed method} 
\label{comparison}

While both the EB and proposed methods minimize the entropy loss subject to moment constraints derived from target covariate summaries, they differ in several key aspects. Table~\ref{tab:comparison} summarizes differences in their covariate shift model parameterization, objective functions, weight interpretations, and resulting estimators.  The objective function of the EB method is used to estimate $\blambda$, while the objective function of the proposed method is used to estimate $(\bmeta,\balpha,\bphi)$. Notably, the EB method estimates weights $\widehat p_i$ defined by \eqref{pi} to align the source and target covariate distributions, whereas the proposed method attains the same objective using the model-based weights $\pi(\bx_i;\halpha)$. 
More importantly, the validity of EB is not guaranteed if $\log \{w(\bx)\}$ is nonlinear in $\bPhi(\bx)$, whereas the proposed method allows a more general specification of $\pi(\bx;\balpha)$ that can capture complex dependencies on $\bx$.

\begin{table}
\caption{Differences between the EB method and the proposed method }
    \begin{tabular}{lcccc}
     \hline
         &&   EB   &   proposed   \\[1ex]  \hline 
     \makecell[c]{Underlying model \\ for $w(\bx)$}   && $ \exp\{a_0+\bm a^\top \bPhi(\bx)\}$ & $\pi(\bx;\balpha)$ \\ [3ex]
      Objective function && $-n\log \sumi \exp[\blambda^\top \{\bPhi(\bx_{i})-\hphi^*\}]$  & \makecell[c]{$ - n\log  \sumi \exp\{\bmeta^\top \bh(\bx_i;\balpha,\bphi)\}$\\ $\;\;\;\;\; +  \frac{m}{2}(\hphi^*-\bphi)^\top \hSigma^{(0)-1} (\hphi^*-\bphi)$}\\[3ex]
      \makecell[l]{Interpretation of \\  estimated weights} && $\widehat p_i$ estimates $n^{-1} w(\bx_i)$ & $\widehat q_i$ estimates $dF(y_i,\bx_i)$ \\[3ex]
      Estimator of $\mu^*$ && $\sumi \hp_i y_i$ & $\sumi \widehat q_i \pi(\bx_i;\halpha)y_i$ \\[3ex] 
    Estimator of $\Sigma^*$ &&  
    $\sumi \widehat p_i \bPhi(\bx_i)^{\otimes 2} -\hphi^{*\otimes 2}$ &   
    $\sumi \widehat q_i\pi(\bx_i;\halpha) \bPhi(\bx_i)^{\otimes 2} -\hphi^{\otimes 2}$\\  \hline
    \end{tabular}
    \label{tab:comparison}
\end{table}

As summarized in Table \ref{tab:comparison}, a key difference between the EB and proposed methods lies in the modeling choice for $w(\bx)$. A natural question is how the EB method relates to the proposed method when both adopt the same covariate shift model. Let $\hmu_{\bPhi}$ denote the proposed estimator calculated by specifying 
$\pi(\bx;\balpha) = \exp\{c+\bm \xi^\top \bPhi(\bx)\}$ with $\balpha=(c,\bm\xi^\top)^\top$. Interestingly,  we can prove that $\hmu_{\bPhi}$ is identical to $\hmu_{\rm EB}$. In this sense, the EB method becomes a special case of the proposed method under this particular model specification. Given that $d_{\balpha}=K+1$, this setting is analogous to the just-identified case considered in \cite{imbens2002generalized}. We can show that $\widehat q_i=n^{-1}$ and $\widehat p_i = n^{-1}\pi(\bx_i;\halpha)$, which leads to the identical weighted estimator. The detailed proof is given in Section 6 of the supplementary materials.

Next, we compare asymptotic efficiency under correct specification of covariate shift, where $\log\{w(\bx)\}$ is linear in $\bPhi(\bx)$ and $w(\bx)$ is correctly specified by $\pi(\bx;\balpha)$. In this case, both the EB and proposed estimators are consistent, but their asymptotic variances may differ depending on the forms of $w(\bx)$ and $\pi(\bx;\balpha)$. First, when $\log\{w(\bx)\}$ is a linear combination of $\bPhi(\bx)$, we have $\hmu_{\rm EB}=\hmu_{\bPhi}$ and thus their asymptotic variances are identical.  Second, consider the case where $\log\{w(\bx)\}$ depends only on a subset $\bPhi_1(\bx)$ of $\bPhi(\bx)$. In this case, we show that $\hmu_{\bPhi_1}$ is asymptotically at least as efficient as $\hmu_{\rm EB}$, where $\hmu_{\bPhi_1}$ is calculated by specifying $\log\{\pi(\bx;\balpha)\}$ to be a linear combination of $\bPhi_1(\bx)$. 
The potential efficiency gain arises because covariate shift modeling and moment constraints can be specified separately under our proposed framework, allowing a correctly specified and parsimonious model while utilizing all available moment information. Consequently, even if one is certain that $\log\{w(\bx)\}$ only involves $\bPhi_1(\bx)$ and constructs an EB estimator using only $\bPhi_1(\bx)$  for moment matching, such an estimator may not achieve the same efficiency as $\hmu_{\bPhi_1}$ because it does not exploit the full set of available moment constraints.  
The comparison is summarized in Theorem \ref{thm4}, with the proof given in Section 7 of the supplementary materials.
\begin{theorem}
\label{thm4} 
Assume that conditions (C1)--(C9) in Section 1 of the supplementary materials are satisfied, $\log\{w(\bx)\}$ is linear in $\bPhi(\bx)$ and $w(\bx)$ is correctly specified by $\pi(\bx;\balpha)$. Then the proposed estimator is asymptotically at least as efficient as the EB estimator. 
\end{theorem}

It is worthwhile to note that, unlike the proposed method, the EB method cannot account for uncertainty in $\widehat{\bm\phi}^*$ directly by augmenting the objective function. To see this, consider minimizing 
$
 \sumi n p_i\log (np_i) +  m (\hphi^*-\bphi)^\top \Sigma^{*-1} (\hphi^*-\bphi)/2 
$
subject to 
$p_i>0,$ $\sumi p_i=1$ and $\sumi p_i \{\bPhi(\bx_i)-\bphi\}=\bm 0.$
The resulting EB estimator $\widetilde\mu_{\rm EB}$ converges to  
$\operatorname{E}  ( \exp[\blambda_{\rm EB}^\top\{\bm\Phi(\bX)-\bphi_{\rm EB}\}]Y )/ \operatorname{E} ( \exp[\blambda_{\rm EB}^\top\{\bm\Phi(\bX)-\bphi_{\rm EB}\}] )$ as $m,n\to \infty$, where $\blambda_{\rm EB}$ and $\bphi_{\rm EB}$ satisfy 
\begin{eqnarray}
\label{ee1}
\operatorname{E} \left( \left\{\bm\Phi(\bX)-\bm \phi_{\rm EB} \right\}   \exp  \left[ \blambda_{\rm EB}^\top   \{\bm\Phi(\bX)-\bm\phi_{\rm EB}\}   \right] \right) =\bm 0,
\end{eqnarray}
\begin{eqnarray}
\label{ee2}
 \blambda_{\rm EB} - \rho^{-1} \Sigma^{*-1} (\bphi^*-\bphi_{\rm EB})=\bm 0.   
\end{eqnarray} 
As shown in Section \ref{method1}, the estimating equation \eqref{eq-lambda} is crucial for proving consistency of $\hmu_{\rm EB}$. Similarly, $\widetilde\mu_{\rm EB}$ is consistent if $\operatorname{E} \left(\{\bPhi(\bX)-\bphi^*\} \exp[\blambda_{\rm EB}^\top \{\bPhi(\bX)-\bphi^*\}] \right) =\bm 0$, which holds when $\bphi_{\rm EB} = \bphi^*$ due to \eqref{ee1}. However, it follows from \eqref{ee1} and \eqref{ee2} that $\bphi_{\rm EB} = \bphi^*$ leads to $\blambda_{\rm EB}=\bm 0$ and $\operatorname{E}\{\bPhi(\bX)\} =\bphi^*$. This implies if $\operatorname{E}\{\bPhi(\bX)\} \neq \bphi^*$, which is commonly encountered under covariate shift, we have $\bphi_{\rm EB} \neq \bphi^*$ and $\operatorname{E} \left(\{\bPhi(\bX)-\bphi^*\} \exp[\blambda_{\rm EB}^\top \{\bPhi(\bX)-\bphi^*\}] \right) =\bm 0$ fails to hold.
We provide an example showing that $\widetilde\mu_{\rm EB}$ is biased when both $\operatorname{E}(Y|\bX)$ and $\log\{w(\bX)\}$ are linear in $\bPhi(\bX)$. Details are given in Section 8 of the supplementary materials.

\section{Numerical simulations}
\label{simulation}
 
Simulations are conducted to evaluate the finite-sample performance of the proposed methods.  To generate data, we exploit the equivalence between model \eqref{x-shift} and a logistic regression for classifying source versus target membership. Let $(Y^\dagger,\bX^\dagger)$ denote the outcome and covariates in a superpopulation consisting of individuals drawn from either the source or the target population.
Define $\delta$ as a binary indicator of population membership, with $\delta=1$ for the target population and $\delta=0$ for the source population. Under covariate shift, let $\Pr^\dagger(\delta=1| Y^\dagger, \bX^\dagger)=\Delta(\bX^\dagger)$, where $\Pr^\dagger$ denotes the probability distribution over the superpopulation. This leads to  $w(\bx)\propto \Delta(\bx)/\{1-\Delta(\bx)\}$.
If data are generated from the joint distribution of $(Y^\dagger,\bX^\dagger)$ and assign membership via $\delta$, the resulting source and target samples follow model \eqref{x-shift}. The covariate $X_1^\dagger$ was generated from the standard normal distribution. Conditional on $X_1^\dagger$, $X_2^\dagger$ and $X_3^\dagger$ were generated by ${\rm  logit} \{\Pr^\dagger(X_2^\dagger=1| X_1^\dagger)\} = -2 X_1^\dagger$ and ${\rm  logit} \{\Pr^\dagger(X_3^\dagger=1| X_1^\dagger)\} = X_1^\dagger$, respectively. Given $\bX^\dagger=(X_1^\dagger,X_2^\dagger,X_3^\dagger)^\top$, we generated the continuous outcome from the normal distribution with mean $g(\bX^\dagger)$ and variance 1,   and generated the binary outcome by ${\rm  logit} \{\Pr^\dagger(Y^\dagger=1| \bX^\dagger)\} = g(\bX^\dagger),$ where $g(\bX^\dagger)= X_1^\dagger +X_2^\dagger +X_3^\dagger-4 X_1^\dagger X_2^\dagger-2$. Consider four scenarios for $\Delta(\bx)$: (i) ${\rm  logit} \{\Delta(\bx)\}= 0.2x_1+0.2x_2+0.2x_3+0.2 x_1^2$; (ii) ${\rm  logit} \{\Delta(\bx)\}= 0.4 x_1$; (iii) ${\rm  logit} \{\Delta(\bx)\}= 0.2x_1+0.2x_2-0.3 x_1x_2$;  (iv) ${\rm  logit} \{\Delta(\bx)\}= 0.2x_1+0.2x_2-0.4 x_1x_2$. 
The target and source sample sizes are $m \in \{250, 500\}$ and $n \in \{m, 2m\}$, respectively.
For the target samples, only covariate summaries $\hphi^*$ are used, which estimates $\bphi^*=(\operatorname{E}^*(X_1^*), \operatorname{E}^*(X_2^*), \operatorname{E}^*(X_3^*), \operatorname{E}^*(X_1^{*2}))^\top$. This corresponds to $\bPhi(\bx)=(x_1,x_2,x_3,x_1^2)^\top$.

Table \ref{tab-s1} and Table \ref{tab-s2} summarize the simulation results for  $\bar y=n^{-1}\sumi y_i$, the EB estimator $\hmu_{\rm EB}$, and the proposed estimator $\hmu$. To obtain $\hmu$, let $\pi(\bx;\balpha)=\exp(\alpha_0+\sum_{j=1}^3 \alpha_j x_j + \alpha_4 x_1^2)$ in scenario (i),  $\pi(\bx;\balpha)=\exp(\alpha_0+ \alpha_1 x_1)$ in scenario (ii), and $\pi(\bx;\balpha)=\exp(\alpha_0+ \alpha_1 x_1 + \alpha_2 x_2 + \alpha_3 x_1 x_2)$ in scenarios (iii) and (iv). We report the empirical bias, the standard deviation of the estimates, the average of the estimated standard errors, and the empirical coverage probability of the 95\% confidence interval based on 1,000 replications. Table \ref{tab-s1} and Table \ref{tab-s2} summarize the simulation results for the continuous outcome and the binary outcome, respectively. In scenarios (i) and (ii) where $\log\{w(\bx)\}$ is linear in $\bPhi(\bx)$, the biases of $\hmu_{\rm EB}$ and $\hmu$ are small, and the coverage probabilities of the 95\% confidence intervals approximate the nominal level. In scenario (i), $\log\{w(\bx)\}$ is a linear combination of $\bPhi(\bx)$, and thus $\hmu_{\rm EB}$ and $\hmu$ are identical. In scenario (ii) where $\log\{w(\bx)\}$ is a linear combination of $x_1$, $\hmu$ enjoys efficiency gains, with the relative efficiency ranging from 1.08 to 1.61. In scenarios (iii) and (iv), where $\log\{w(\bx)\}$ is not linear in $\bPhi(\bx)$, $\hmu_{\rm EB}$ yields larger biases, which do not decrease as the sample size increases. Moreover, $\hmu_{\rm EB}$ does not achieve satisfactory coverage probabilities. From scenario (iii) to (iv), a larger contribution from the interaction $x_1x_2$ to covariate shift increases the biases of $\hmu_{\rm EB}$. By correctly specifying the covariate shift model, $\hmu$ yields smaller biases and maintains coverage probabilities close to 95\%. In all scenarios, the biases of $\bar y$ are large due to ignoring covariate shift.

\begin{table}
	\caption{\label{tab-s1} Summary of simulation results when the outcome is continuous}
    \begin{threeparttable}
\setlength{\tabcolsep}{1.5 mm}
\begin{tabular}{cccccccccccccccccc}
 \hline
                     &       && \multicolumn{4}{c}{$\bar y$}     && \multicolumn{4}{c}{ $ \hmu_{\rm EB}$}     && \multicolumn{4}{c}{ $ \hmu$}             \\ \cline{4-7} \cline{9-12}  \cline{14-17}  
$n$    & $m$  && BIAS     & SD   & SE &  CP &&  BIAS    & SD   & SE &  CP  &&  BIAS    & SD   & SE &  CP \\ \hline
\multicolumn{16}{c}{Scenario (i)}  \\ [1ex]
500 & 250 && -199 &	98	& 99 & 47.7  && ~4 & 139	& 140 & 94.5   && ~4 & 139	& 140 & 94.5   \\[0.5ex]
500 & 500 && -199 &	98	& 99 & 47.7  && ~3 & 112	& 113 & 95.1   &&  ~3 & 112	& 113 & 95.1   \\[0.5ex]
1000 & 500 && -197	& 70 & 70 & 19.4   && -2 & ~96 & 100	& 95.2   &&  -2 & ~96 & 100	& 95.2 \\[0.5ex]
1000 & 1000 && -197 & 70  & 70 & 19.4   && ~2	& ~76 & ~81 & 95.9  && ~2	& ~76 & ~81 & 95.9  \\[1ex]

\multicolumn{16}{c}{Scenario (ii)}  \\ [1ex]
500 & 250 && 420 & 110 & 108 & 2.9  && 4 & 120 & 118 & 94.6   && 7 & 98	& 98 & 95.4   \\[0.5ex]  
500 & 500 && 420 & 110 & 108 & 2.9  && 3 & ~99	& ~97 & 95.1  && 6 & 89	& 87 &	94.9 \\[0.5ex]  
1000 & 500 && 427 & ~77 & ~76 & ~0.0  && 3 & ~87 & ~83 & 93.2   && 6 & 69	& 69 & 94.3  \\[0.5ex] 
1000 & 1000 && 427 & ~77 & ~76 & ~0.0  && 5	& ~70 & ~69 & 94.2  && 6	& 62 & 61 & 94.9  \\[1ex] 

\multicolumn{16}{c}{Scenario (iii)}  \\ [1ex]
500 & 250 && -232 & 99	& 100 & 37.1  && -88 & 133 & 132 & 88.6  && -13	& 175 &	171	& 95.2  \\[0.5ex]
500 & 500 && -232 & 99	& 100 & 37.1  && -86 & 106 & 105 & 86.6 && ~-7 & 135	& 132	& 94.2  \\[0.5ex]
1000 & 500 && -229	& 72 & ~71 & 11.1  && -86 & ~96 & ~93 & 82.1 && ~-5 & 123 & 120 & 94.7   \\[0.5ex]
1000 & 1000 &&  -229 & 72 & ~71 & 11.1  && -84 & ~74 & ~74	& 77.4  && ~~0 & ~91 & ~93 &	94.4  \\[1ex]

\multicolumn{16}{c}{Scenario (iv)}  \\ [1ex]
500 & 250 && -347 & 97 & 98	& 5.8 && -116 & 137	& 134 & 84.3 && -11	& 178 & 172	 & 93.5  \\[0.5ex]
500 & 500 && -347 & 97 & 98	& 5.8  &&  -115	& 108 & 107	& 80.9 && ~-7 & 136 & 133	& 94.0   \\[0.5ex]
1000 & 500 && -344	& 71 & 69 & 0.5  && -115 & ~99 & ~95	& 73.6 && ~-4 & 125 & 120	& 94.2 \\[0.5ex]
1000 & 1000 && -344	& 71 & 69 & 0.5  && -113 & ~76 & ~75	& 65.7 && ~~0	& ~92 & ~93	& 95.0   \\

\hline
\end{tabular} 
\begin{tablenotes}
     \item[] {\footnotesize  NOTE: BIAS, empirical bias ($\times 1000$); SD, standard deviation ($\times 1000$); SE, average of the estimated standard errors ($\times 1000$); CP, empirical coverage probability (\%) at the 95\% confidence level.}
  \end{tablenotes}
  \end{threeparttable}
\end{table}

\begin{table}
	\caption{\label{tab-s2} Summary of simulation results when the outcome is binary}
    \begin{threeparttable}
\setlength{\tabcolsep}{1.5 mm}
\begin{tabular}{cccccccccccccccccc}
\hline
                     &       && \multicolumn{4}{c}{$\bar y$}     && \multicolumn{4}{c}{ $ \hmu_{\rm EB}$}     && \multicolumn{4}{c}{ $ \hmu$}             \\ \cline{4-7} \cline{9-12}  \cline{14-17}  
$n$    & $m$  && BIAS     & SD   & SE &  CP &&  BIAS    & SD   & SE &  CP  &&  BIAS    & SD   & SE &  CP \\ \hline
\multicolumn{16}{c}{Scenario (i)}  \\ [1ex]
500 & 250 && -30 & 22 & 22 & 72.0  && ~1	& 27 & 28 & 95.0 && ~1	& 27 & 28 & 95.0  \\[0.5ex]
500 & 500 && -30 & 22 & 22 & 72.0  && ~1	& 24 & 26 & 96.1 && ~1	& 24 & 26 & 96.1  \\[0.5ex]
1000 & 500 && -31 & 16 & 16	& 48.9 && -1 & 19 & 20 & 95.1  && -1 & 19 & 20 & 95.1   \\[0.5ex]
1000 & 1000 && -31 & 16  & 16 &  48.9 && ~0 & 18	& 18 & 94.7  &&  ~0 & 18	& 18 & 94.7     \\[1ex]

\multicolumn{16}{c}{Scenario (ii)}  \\ [1ex]
500 & 250 && 51	& 23 & 22 & 39.8 &&  ~0	& 27 & 26 & 94.4 && 1 & 25 & 24	& 94.7   \\[0.5ex] 
500 & 500 &&  51 & 23 & 22 & 39.8 && ~0	& 25 & 24 & 94.7 &&  1	& 24 & 23 & 94.9  \\[0.5ex] 
1000 & 500 && 51 & 16 & 16 & 10.2 && -1	& 20 & 19 & 93.7  && 0	& 18 & 17 & 94.5  \\[0.5ex] 
1000 & 1000 && 51 &  16  & 16 & 10.2 && ~0 & 18 & 17 & 94.5 && 0  & 17  & 17    &   94.7 \\[1ex]

\multicolumn{16}{c}{Scenario (iii)}  \\ [1ex]
500 & 250 && -36 & 22  & 22 & 64.5  && -14	& 25 & 25 & 90.4  && -1	& 31 & 30 & 95.2  \\[0.5ex]
500 & 500 &&  -36 & 22  & 22 & 64.5  && -14	& 23 & 23 & 89.8 && -1 & 27	& 26 & 94.6   \\[0.5ex]
1000 & 500 && -36 & 16 & 16 & 35.7   && -15	& 19 & 18 & 84.5 && -1 & 22	& 21 &	94.1  \\[0.5ex]
1000 & 1000 && -36 & 16 & 16 & 35.7  && -14	& 17 & 16 & 83.8  && ~0 & 19	& 18 &	93.7  \\[1ex]

\multicolumn{16}{c}{Scenario (iv) }  \\ [1ex]
500 & 250 && -51 & 22  & 22 & 39.1   && -18	& 26 & 25 & 88.0 && -1  & 32  &  30   & 94.6  \\[0.5ex]
500 & 500 &&  -51 & 22  & 22 & 39.1  && -18	& 23 & 23 & 85.8 && ~0  & 27  & 26    & 94.0   \\[0.5ex]
1000 & 500 && -52 & 16  & 16 & 10.3   && -19 & 19 & 18 & 78.8 && -1  & 22  & 21    & 94.4   \\[0.5ex]
1000 & 1000 && -52 & 16  & 16 & 10.3  && -19 & 17 & 16 & 76.3 && ~0 & 19  &  18   & 94.6   \\

\hline
\end{tabular} 
\begin{tablenotes}
     \item[] {\footnotesize  NOTE: BIAS, empirical bias ($\times 1000$); SD, standard deviation ($\times 1000$); SE, average of the estimated standard errors ($\times 1000$); CP, empirical coverage probability (\%) at the 95\% confidence level.}
  \end{tablenotes}
  \end{threeparttable}
\end{table}

We conducted simulations to evaluate the Type I error rate of the model checking procedure in Section 10 of the supplementary materials. Simulations under model misspecification are reported in Section 12 of the supplementary materials. 
Table 5 in the supplementary materials shows that, under model misspecification, the proposed estimator exhibits some bias, the average estimated standard errors deviate from the standard deviations, and the resulting confidence intervals fail to attain the nominal coverage probability. Nevertheless, the proposed and EB estimators exhibit comparable biases and standard deviations.

\vspace{-0.2 in}

\section{Data application}
\label{realdata}

Breast cancer is the most frequently diagnosed cancer globally and the leading cause of cancer-related death among women. The Surveillance, Epidemiology, and End Results (SEER) Program of the National Cancer Institute (NCI) is a cancer surveillance program covering approximately 45.9\% of the U.S. population. 
The SEER Program collects and publishes data on cancer incidence, survival, mortality, patient demographics, and tumor characteristics.  We illustrated the EB and proposed methods using SEER breast cancer data.

Existing research demonstrates disparities in survival outcomes across populations with different incomes \citep{singh2003area}. It has been recognized that lower-income patients may be less likely to participate in clinical trials \citep{unger2013patient}, and analyses based on such trials often fail to fully reflect these disparities. In contrast, the SEER Program consists of patients from diverse socioeconomic backgrounds, thereby providing a valuable opportunity to evaluate disparities in mortality probabilities across populations with different incomes. To this end, we treated the breast cancer patients diagnosed in 2015 with a median household income below $40,000$ dollars as the target population.  In the target population, let $T^*$ be the survival time in months since cancer diagnosis. The outcome was denoted by $Y^*(t)=\operatorname{I}(T^* < t)$. The parameter of interest was defined as the mortality at $t$ months, that is, $\mu^*(t)=\operatorname{E}^*\{Y^*(t)\}$ for $t \in \{1,\ldots,24\}$.
The source population consisted of patients diagnosed in 2010 who had a median household income exceeding $75,000$ dollars and resided in metropolitan areas with populations ranging from 250,000 to 1 million. In the source population, let $T$ be the survival time in months and let $Y(t)=\operatorname{I}(T < t)$ be the outcome. Consider six baseline covariates: tumor size, age at diagnosis, stage (SEER Summary), estrogen receptor (ER), human epidermal growth factor receptor 2 (HER2), and progesterone receptor (PR). Existing research demonstrated that distributions of covariates differ between the two populations with different incomes \citep{Miller2002impact}.

\begin{figure}
    \subfigure[$\bar y(t)$]{\includegraphics[width=5.4cm]{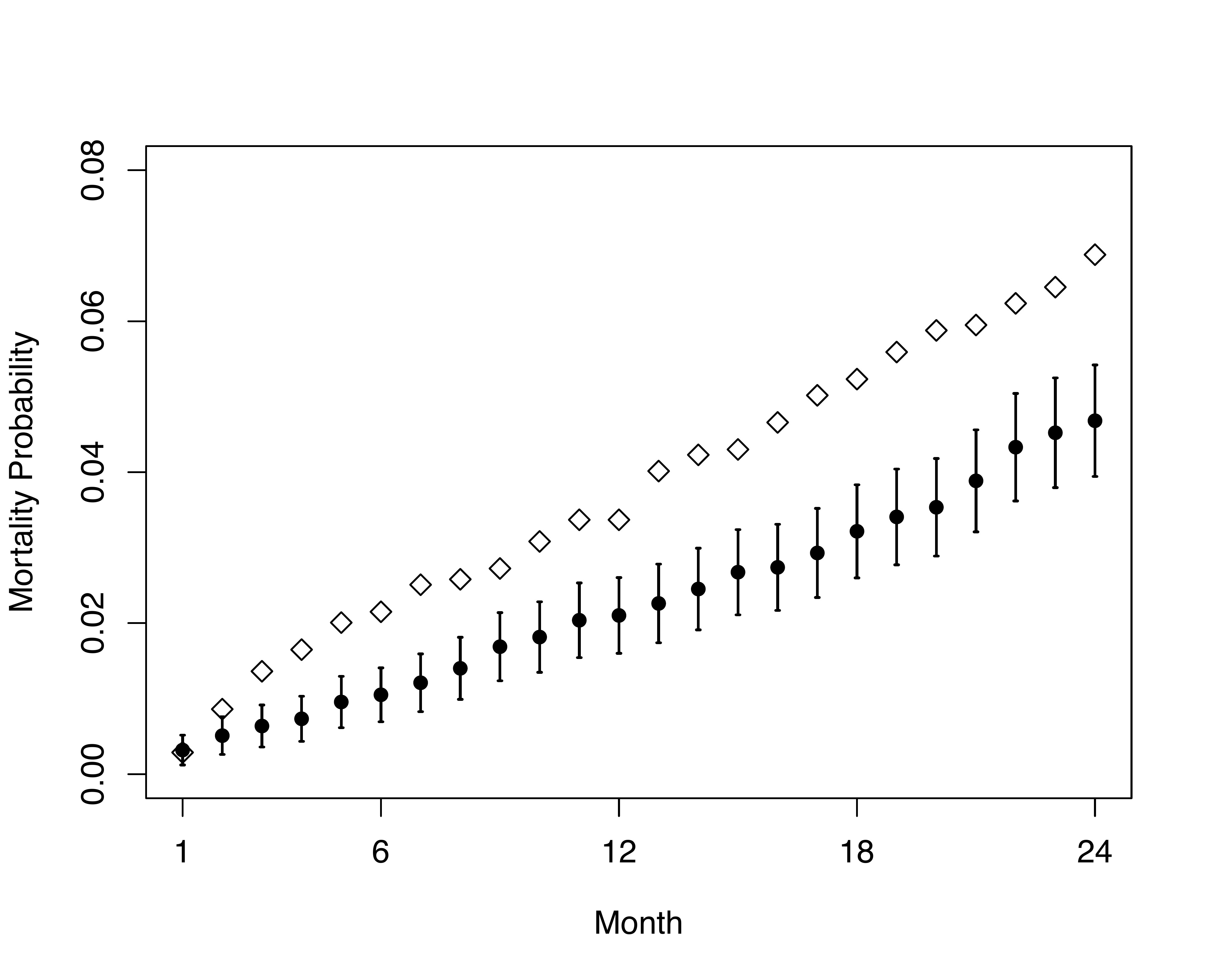}}
	\subfigure[$\hmu_{\rm EB}(t)$]{\includegraphics[width=5.4cm]{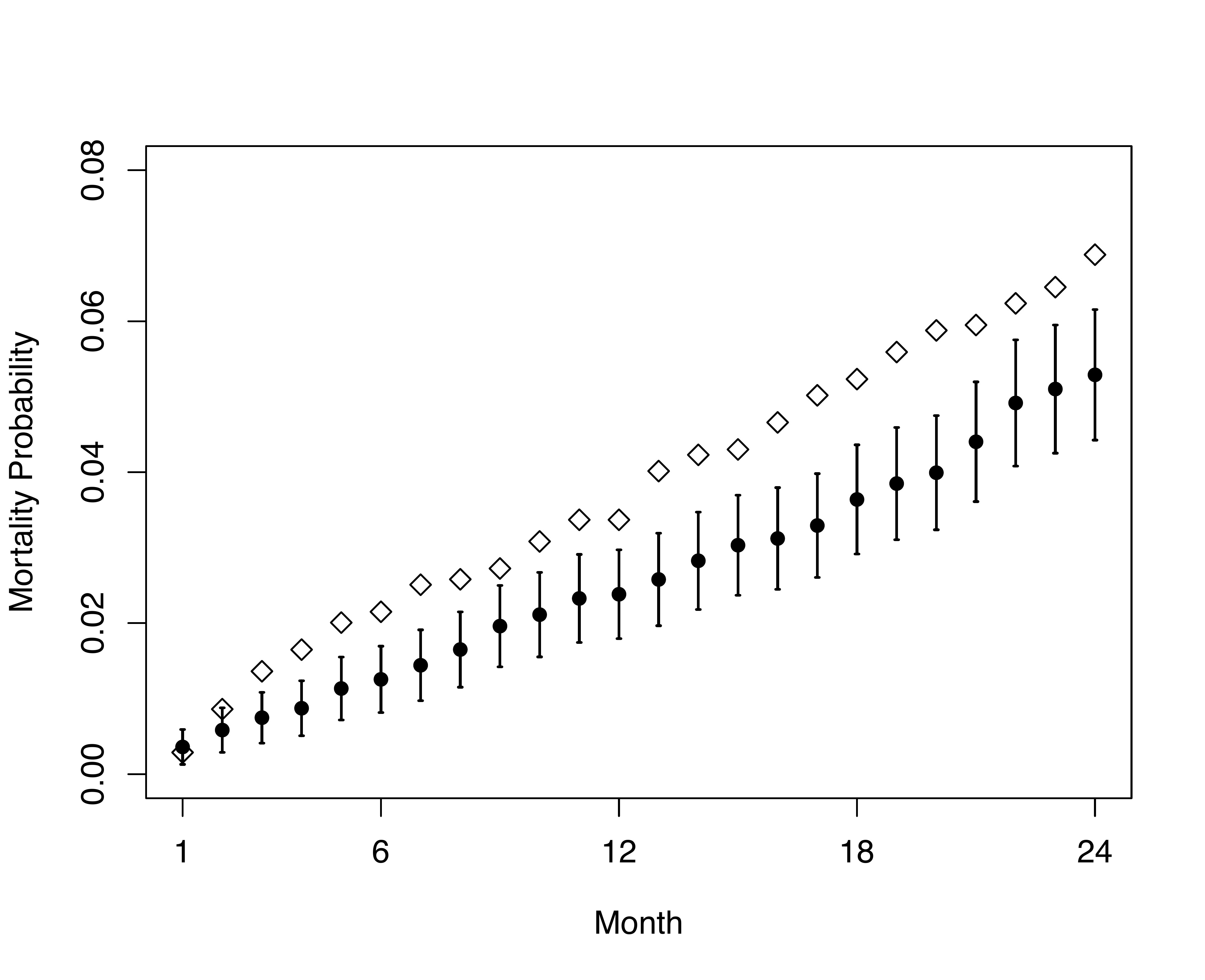}}
    \subfigure[$\hmu(t)$]{\includegraphics[width=5.4cm]{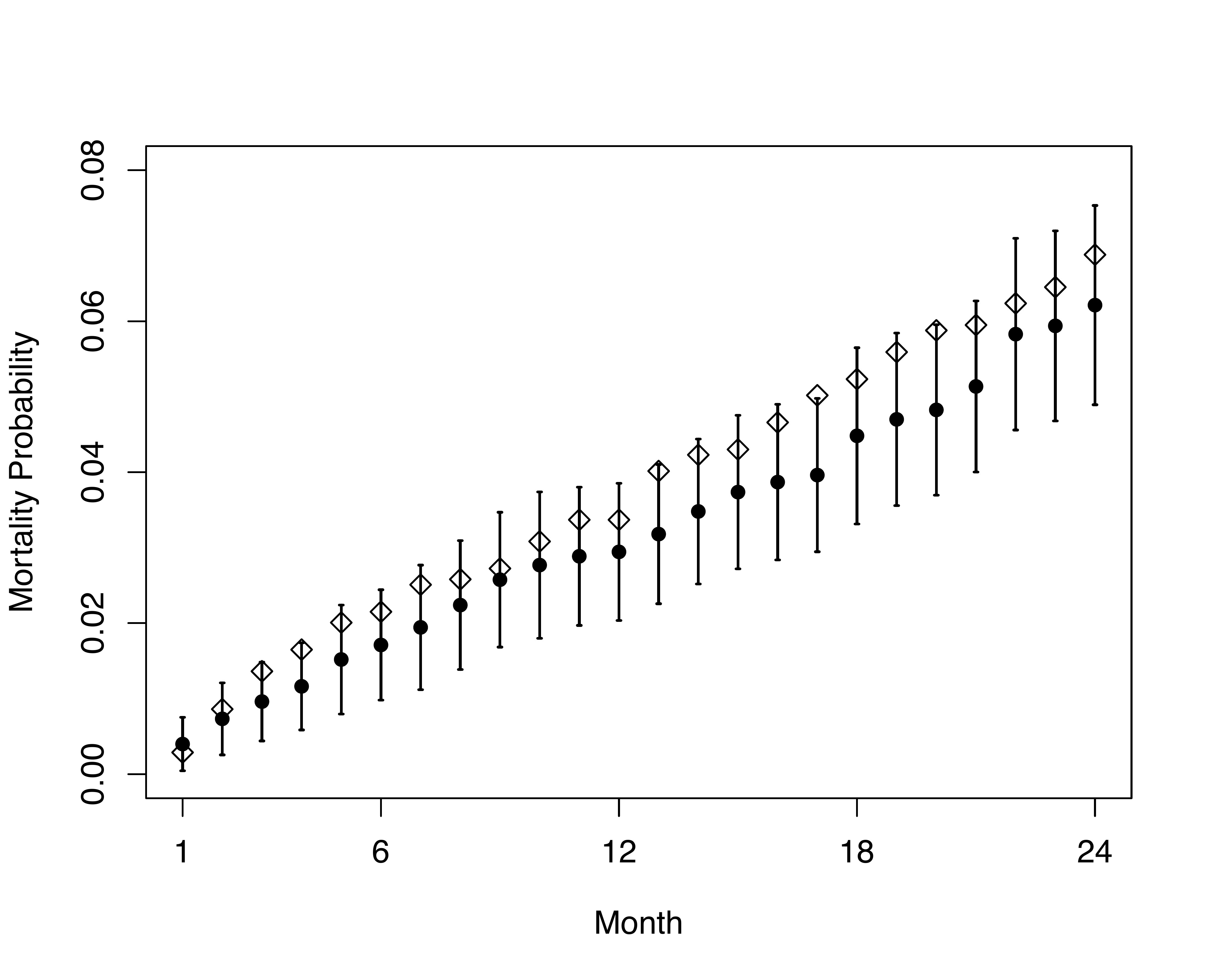}}
    \caption{The mortality probabilities of $\bar y^*(t)$ ($\diamond$) and the estimated mortality probabilities ($\bullet$) with the corresponding 95\% confidence intervals (vertical bars) for $\bar y(t)$, the EB estimator $\hmu_{\rm EB}(t)$, and the proposed estimator $\hmu(t)$ at the $t-$months, $t \in \{1,\ldots,24\}$.  }\label{fig:mortality}
\end{figure}

The target study consisted of $m=1,395$ patients. For $j=1,\ldots,m$, let $t_j^*$ denote the observed survival time in the target study and define $y_j^*(t)=\operatorname{I}(t_j^* < t)$. For $t \in \{1,\ldots,24\}$, $y_j^*(t)$ was not censored and we used  $\bar y^*(t)=m^{-1} \sum_{j=1}^m y_j^*(t)$ as a benchmark to evaluate the performance of the proposed methods. The source study consisted of $n=3,140$ patients. For $i=1,\ldots,n$, let $t_i$ denote the observed survival time in the source study and define $y_i(t)=\operatorname{I}(t_i < t)$. For $t \in \{1,\ldots,24\}$, $y_i(t)$ was not censored and we calculated $\bar y(t)=n^{-1} \sumi y_i(t)$ with the 95\% confidence intervals.  As shown in Figure \ref{fig:mortality} (a), there is a significant difference between $\bar y^*(t)$ and $\bar y(t)$. 
Hence, we apply the proposed methods to correct the bias of $\bar y(t)$.

We first applied EB to estimate $\mu^*(t)$. 
Denote the covariates in the source population as tumor size in millimeters ($X_1$), age ($\ge $ 60 years vs $<60$ years, $X_2$), stage (localized vs non-localized, $X_3$), ER (positive vs negative, $X_4$), HER2 (positive vs negative, $X_5$), and PR (positive vs negative, $X_6$). The covariate summaries from the source and target populations are summarized in Table \ref{tab:summary}. 
We calculated the EB weights using the source individual-level covariate data and the target covariate summaries provided in Table \ref{tab:summary}, where the second-order moment of tumor size from the target population was derived from its mean and variance. Given the EB weights, we calculated the EB estimator $\hmu_{\rm EB}(t)$ and the corresponding 95\% confidence interval.  As shown in Figure \ref{fig:mortality} (b), despite $\hmu_{\rm EB}(t)$ being closer to $\bar y^*(t)$ than $\bar y(t)$, there remains a discrepancy between $\hmu_{\rm EB}(t)$ and $\bar y^*(t)$. This indicates that $\log \{w(\bx)\}$ may not be linear in $\bPhi(\bx)=(\bx^\top,x_1^2)^\top$, where $\bx=(x_1,\ldots,x_6)^\top$.

We then applied the proposed method by imposing an appropriate model $\pi(\bx;\balpha)$. Existing research demonstrated distributional shifts of tumor size, stage, and ER between two populations with different incomes \citep{Miller2002impact, krieger2016metrics}. Table \ref{tab:summary} showed that the distributions of tumor size, age, stage, and ER were different between these two populations. We therefore initially included $(x_1,x_2,x_3,x_4,x_1^2)$ in $\pi(\bx;\balpha)$. To capture potential interactions, we considered three candidate models by incorporating $x_1x_2$, $x_1 x_3$ and $x_1 x_4$, respectively. Each candidate model was assessed using the model checking procedure. The inclusion of $x_1x_3$ and $x_1 x_4$ yielded p-values less than 0.01, leading to rejection of the corresponding models. In contrast, the inclusion of $x_1x_2$ yielded a p-value of 0.26, providing no compelling evidence to reject the inclusion of $x_1x_2$.
Therefore, we specified $\pi(\bx;\balpha)=\exp(\alpha_0+\sum_{j=1}^4 \alpha_j x_j + \alpha_5 x_1^2 + \alpha_6 x_1 x_2)$ and calculated $\hmu(t)$ with its 95\% confidence interval. Figure \ref{fig:mortality} (c) displays $\hmu(t)$ along with the corresponding 95\% confidence interval. As shown in the figure, $\hmu(t)$ is close to $\bar y^*(t)$, and $\bar y^*(t)$ falls within the 95\% confidence interval. Additional data analyses, including assessment of covariate shift and numerical results corresponding to Figure \ref{fig:mortality}, are provided in Section 13 of the supplementary materials.

\begin{table}
	\caption{\label{tab:summary} Covariate summary statistics from the source and target populations}
    \centering
	\begin{threeparttable}
\begin{tabular}{ccc}
 \hline
                     &  Source ($n=3,140$)        &     Target ($m=1,395$)  \\[1ex]       
                     
 \hline 
tumor size in millimeters (mean)  &  17.9 &  18.9 \\ 
tumor size in millimeters (variance) & 124.4 & 129.5 \\  
age at diagnosis $\ge $ 60 years  & 54.6\% & 64.5\%\\ 
localized stage  & 69.7\%  & 68.2\%\\  
ER positive  & 84.3\% & 82.9\% \\     
HER2 positive & 14.2\% & 14.5\% \\ 
PR  positive & 73.9\%  & 74.2\%   \\           
\hline
\end{tabular}
\begin{tablenotes}
     \item[] {\footnotesize  NOTE: ER, estrogen receptor; HER2, human epidermal growth factor receptor 2; PR, progesterone receptor.}
  \end{tablenotes}
  \end{threeparttable}
\end{table}

\vspace{-0.5 in}

\section{Discussion}
This paper presents a new transportability method for estimation and inference in the target population without the target individual-level data. 
Throughout this paper, we assume that the conditional distributions of the outcome given covariates are the same between the source and target populations, which is stronger than the mean exchangeability condition  \citep{chen2023entropy, chen2024robust, chen2026confidence}. We adopt this assumption because we aim to develop a unified transportability framework that can transport general parameters, not only $\operatorname{E}^*(Y^*)$.  We prove in Section 9 of the supplementary materials that $\halpha$ is consistent for $\balpha_0$ and asymptotically normal. Given  $\widehat q_i$ and $\pi(\bx;\halpha)$, the proposed method can be extended to estimate a general parameter of interest in the target population,  which is presented in Section 11 of the supplementary materials. 
In practice, convergence failure may be attributable to several factors, including violation of the positivity assumption, nonconvexity of the optimization problem, or misspecification of $\pi(\bx;\balpha)$. Violation of positivity arises when the source covariate support does not adequately cover the target covariate support, which can be assessed based on prior information on study designs and available covariate summaries. Moreover, we recommend employing multiple initial values to diagnose numerical instability due to nonconvexity and using the proposed model checking procedure to assess model specification.

\backmatter


\section*{Acknowledgements}
This work was supported by the National Natural Science Foundation of China No. 12201616 (Sheng), and National Institutes of Health grants R01HL173153 (Sun, Huang), and P01AG082653 (Huang). 
The authors acknowledge the efforts of the NCI and the SEER Program tumor registries. The authors are solely responsible for the interpretation and reporting of the SEER data. The ideas and opinions expressed herein are those of the author(s) and do not necessarily reflect the opinions of NCI.

\section*{Supplementary materials}
Web Appendices, Tables, and Figures referenced in Sections 3--8 and \texttt{R} code for reproducing the results are available with this paper at the \textit{Biometrics} website on Oxford Academic.

\section*{Data availability}
The Surveillance, Epidemiology, and End Results breast cancer data that support the findings in this paper are available at \texttt{https://seer.cancer.gov/data/access.html}.

\bibliographystyle{biom} 
\bibliography{ref1.bib}

@article{chattopadhyay2024one,
  title={One-step weighting to generalize and transport treatment effect estimates to a target population},
  author={Chattopadhyay, Ambarish and Cohn, Eric R and Zubizarreta, Jos{\'e} R},
  journal={The American Statistician},
  volume={78},
  number={3},
  pages={280--289},
  year={2024},
  publisher={Taylor \& Francis}
}

@article{chen2026confidence,
  title={Confidence Interval Construction for Causally Generalized Estimates With Target Sample Summary Information},
  author={Chen, Yi and Chen, Guanhua and Yu, Menggang},
  journal={Statistics in Medicine},
  volume={45},
  number={1-2},
  pages={e70358},
  year={2026},
  publisher={Wiley Online Library}
}

@article{imbens2002generalized,
  title={Generalized method of moments and empirical likelihood},
  author={Imbens, Guido W},
  journal={Journal of Business \& Economic Statistics},
  volume={20},
  number={4},
  pages={493--506},
  year={2002},
  publisher={Taylor \& Francis}
}

@article{unger2013patient,
  title={Patient income level and cancer clinical trial participation},
  author={Unger, Joseph M and Hershman, Dawn L and Albain, Kathy S and Moinpour, Carol M and Petersen, Judith A and Burg, Kenda and Crowley, John J},
  journal={Journal of Clinical Oncology},
  volume={31},
  number={5},
  pages={536--542},
  year={2013},
  publisher={American Society of Clinical Oncology}
}

@book{singh2003area,
  title        = {Area socioeconomic variations in U.S. cancer incidence, mortality, stage, treatment, and survival, 1975--1999},
  author       = {Singh, Gopal K. and Miller, Barry A. and Hankey, Benjamin F. and Edwards, Brenda K.},
  series       = {NCI Cancer Surveillance Monograph Series},
  number       = {4},
  year         = {2003},
  publisher    = {National Cancer Institute},
  address      = {Bethesda, MD},
  note         = {NIH Publication No. 03-0000}
}

@article{Prentice1979logistic,
    author = {Prentice, R. L. and Pyke, R.},
    title = {Logistic disease incidence models and case-control studies},
    journal = {Biometrika},
    volume = {66},
    number = {3},
    pages = {403--411},
    year = {1979}
}

@article{dahabreh2023efficient,
  title={Efficient and robust methods for causally interpretable meta-analysis: Transporting inferences from multiple randomized trials to a target population},
  author={Dahabreh, Issa J and Robertson, Sarah E and Petito, Lucia C and Hern{\'a}n, Miguel A and Steingrimsson, Jon A},
  journal={Biometrics},
  volume={79},
  number={2},
  pages={1057--1072},
  year={2023},
  publisher={Wiley Online Library}
}

@article{rudolph2017robust,
  title={Robust estimation of encouragement design intervention effects transported across sites},
  author={Rudolph, Kara E and Laan, Mark J},
  journal={Journal of the Royal Statistical Society Series B: Statistical Methodology},
  volume={79},
  number={5},
  pages={1509--1525},
  year={2017},
  publisher={Oxford University Press}
}

@article{colnet2024causal,
  title={Causal inference methods for combining randomized trials and observational studies: a review},
  author={Colnet, B{\'e}n{\'e}dicte and Mayer, Imke and Chen, Guanhua and Dieng, Awa and Li, Ruohong and Varoquaux, Ga{\"e}l and Vert, Jean-Philippe and Josse, Julie and Yang, Shu},
  journal={Statistical Science},
  volume={39},
  number={1},
  pages={165--191},
  year={2024},
  publisher={Institute of Mathematical Statistics}
}

@article{dahabreh2019generalizing,
  title={Generalizing causal inferences from individuals in randomized trials to all trial-eligible individuals},
  author={Dahabreh, Issa J and Robertson, Sarah E and Tchetgen, Eric J and Stuart, Elizabeth A and Hern{\'a}n, Miguel A},
  journal={Biometrics},
  volume={75},
  number={2},
  pages={685--694},
  year={2019},
  publisher={Oxford University Press}
}

@article{josey2022calibration,
  title={A calibration approach to transportability and data-fusion with observational data},
  author={Josey, Kevin P and Yang, Fan and Ghosh, Debashis and Raghavan, Sridharan},
  journal={Statistics in Medicine},
  volume={41},
  number={23},
  pages={4511--4531},
  year={2022},
  publisher={Wiley Online Library}
}

@article{dahabreh2020extending,
  title={Extending inferences from a randomized trial to a new target population},
  author={Dahabreh, Issa J and Robertson, Sarah E and Steingrimsson, Jon A and Stuart, Elizabeth A and Hern{\'a}n, Miguel A},
  journal={Statistics in Medicine},
  volume={39},
  number={14},
  pages={1999--2014},
  year={2020},
  publisher={Wiley Online Library}
}

@article{josey2021transporting,
  title={Transporting experimental results with entropy balancing},
  author={Josey, Kevin P and Berkowitz, Seth A and Ghosh, Debashis and Raghavan, Sridharan},
  journal={Statistics in Medicine},
  volume={40},
  number={19},
  pages={4310--4326},
  year={2021},
  publisher={Wiley Online Library}
}

@article{kitamura1997information,
  title={An information-theoretic alternative to generalized method of moments estimation},
  author={Kitamura, Yuichi and Stutzer, Michael},
  journal={Econometrica},
   number = {4},
  pages={861--874},
  volume = {65},
  year={1997},
  publisher={JSTOR}
}

@article{Imbens1998Information,
title = {Information theoretic approaches to inference in moment condition models},
 author = {Guido W. Imbens and Richard H. Spady and Phillip Johnson},
 journal = {Econometrica},
 number = {2},
 pages = {333--357},
 volume = {66},
 year = {1998},
 publisher = {[Wiley, Econometric Society]},
}

@article{cressie1984multinomial,
  title={Multinomial goodness-of-fit tests},
  author={Cressie, Noel and Read, Timothy RC},
  journal={Journal of the Royal Statistical Society Series B: Statistical Methodology},
  volume={46},
  number={3},
  pages={440--464},
  year={1984},
  publisher={Oxford University Press}
}

@article{wu2023transfer,
  title={Transfer learning of individualized treatment rules from experimental to real-world data},
  author={Wu, Lili and Yang, Shu},
  journal={Journal of Computational and Graphical Statistics},
  volume={32},
  number={3},
  pages={1036--1045},
  year={2023},
  publisher={Taylor \& Francis}
}

@article{degtiar2023review,
  title={A review of generalizability and transportability},
  author={Degtiar, Irina and Rose, Sherri},
  journal={Annual Review of Statistics and Its Application},
  volume={10},
  number={1},
  pages={501--524},
  year={2023},
  publisher={Annual Reviews}
}

@article{qin2008efficient,
  title={Efficient and doubly robust imputation for covariate-dependent missing responses},
  author={Qin, Jing and Shao, Jun and Zhang, Biao},
  journal={Journal of the American Statistical Association},
  volume={103},
  number={482},
  pages={797--810},
  year={2008},
  publisher={Taylor \& Francis}
}

@article{lee2023improving,
  title={Improving trial generalizability using observational studies},
  author={Lee, Dasom and Yang, Shu and Dong, Lin and Wang, Xiaofei and Zeng, Donglin and Cai, Jianwen},
  journal={Biometrics},
  volume={79},
  number={2},
  pages={1213--1225},
  year={2023},
  publisher={Oxford University Press}
}

@article{imai2014covariate,
  title={Covariate balancing propensity score},
  author={Imai, Kosuke and Ratkovic, Marc},
  journal={Journal of the Royal Statistical Society Series B: Statistical Methodology},
  volume={76},
  number={1},
  pages={243--263},
  year={2014},
  publisher={Oxford University Press}
}

@article{han2013estimation,
  title={Estimation with missing data: Beyond double robustness},
  author={Han, Peisong and Wang, Lu},
  journal={Biometrika},
  volume={100},
  number={2},
  pages={417--430},
  year={2013},
  publisher={Oxford University Press}
}

@article{han2014multiply,
  title={Multiply robust estimation in regression analysis with missing data},
  author={Han, Peisong},
  journal={Journal of the American Statistical Association},
  volume={109},
  number={507},
  pages={1159--1173},
  year={2014},
  publisher={Taylor \& Francis}
}

@article{chan2014oracle,
  title={Oracle, multiple robust and multipurpose calibration in a missing response problem},
  author={Chan, Kwun Chuen Gary and Yam, Sheung Chi Phillip},
  journal = {Statistical Science},
  volume={29},
  number={3},
  pages={380--396},
  year={2014}
}

@article{chen2024robust,
  title={Robust sample weighting to facilitate individualized treatment rule learning for a target population},
  author={Chen, Rui and Huling, Jared D and Chen, Guanhua and Yu, Menggang},
  journal={Biometrika},
  volume={111},
  number={1},
  pages={309--329},
  year={2024},
  publisher={Oxford University Press}
}

@article{huang2006correcting,
  title={Correcting sample selection bias by unlabeled data},
  author={Huang, Jiayuan and Smola, Alexander J. and Gretton, Arthur and Borgwardt, Karsten M. and Sch{\"o}lkopf, Bernhard},
  journal={Advances in Neural Information Processing Systems},
  volume={20},
  pages={601-608},
  year={2007}
}

@article{Miller2002impact,
    author = {Miller, Barry A. and Hankey, Benjamin F. and Thomas, Terry L.},
    title = {Impact of sociodemographic factors, hormone receptor status, and tumor grade on ethnic differences in tumor stage and size for breast cancer in {US} women},
    journal = {American Journal of Epidemiology},
    volume = {155},
    number = {6},
    pages = {534-545},
    year = {2002}
}

@article{chan2016globally,
  title={Globally efficient non-parametric inference of average treatment effects by empirical balancing calibration weighting},
  author={Chan, Kwun Chuen Gary and Yam, Sheung Chi Phillip and Zhang, Zheng},
  journal={Journal of the Royal Statistical Society Series B: Statistical Methodology},
  volume={78},
  number={3},
  pages={673--700},
  year={2016},
  publisher={Oxford University Press}
}

@article{chen2023entropy,
  title={Entropy balancing for causal generalization with target sample summary information},
  author={Chen, Rui and Chen, Guanhua and Yu, Menggang},
  journal={Biometrics},
  volume={79},
  number={4},
  pages={3179--3190},
  year={2023},
  publisher={Wiley Online Library}
}

@article{chu2023targeted,
  title={Targeted optimal treatment regime learning using summary statistics},
  author={Chu, Jianing and Lu, Wenbin and Yang, Shu},
  journal={Biometrika},
  volume={110},
  number={4},
  pages={913--931},
  year={2023},
  publisher={Oxford University Press}
}

@article{hainmueller2012entropy,
  title={Entropy balancing for causal effects: A multivariate reweighting method to produce balanced samples in observational studies},
  author={Hainmueller, Jens},
  journal={Political Analysis},
  volume={20},
  number={1},
  pages={25--46},
  year={2012},
  publisher={Cambridge University Press}
}

@article{kanamori2009least,
  title={A least-squares approach to direct importance estimation},
  author={Kanamori, Takafumi and Hido, Shohei and Sugiyama, Masashi},
  journal={Journal of Machine Learning Research},
  volume={10},
  pages={1391--1445},
  year={2009},
  publisher={JMLR. org}
}

@article{krieger2016metrics,
  title={Metrics for monitoring cancer inequities: residential segregation, the Index of Concentration at the Extremes ({ICE}), and breast cancer estrogen receptor status ({USA}, 1992--2012)},
  author={Krieger, Nancy and Singh, Nakul and Waterman, Pamela D},
  journal={Cancer Causes \& Control},
  volume={27},
  pages={1139--1151},
  year={2016},
  publisher={Springer}
}

@article{zheng2022risk,
  title={Risk projection for time-to-event outcome leveraging summary statistics with source individual-level data},
  author={Zheng, Jiayin and Zheng, Yingye and Hsu, Li},
  journal={Journal of the American Statistical Association},
  volume={117},
  number={540},
  pages={2043--2055},
  year={2022},
  publisher={Taylor \& Francis}
}

@article{zhang2020generalized,
  title={Generalized integration model for improved statistical inference by leveraging external summary data},
  author={Zhang, Han and Deng, Lu and Schiffman, Mark and Qin, Jing and Yu, Kai},
  journal={Biometrika},
  volume={107},
  number={3},
  pages={689--703},
  year={2020},
  publisher={Oxford University Press}
}

@article{zhao2017entropy,
  title={Entropy balancing is doubly robust},
  author={Zhao, Qingyuan and Percival, Daniel},
  journal={Journal of Causal Inference},
  volume={5},
  number={1},
  pages={20160010},
  year={2017},
  publisher={De Gruyter}
}


\label{lastpage}

\end{document}